\documentclass[a4paper,10pt,superscriptaddress,twocolumn,pra]{revtex4-2}

\usepackage{amsmath}            
\usepackage{amssymb}            
\usepackage{amsfonts}           
\usepackage{bm}                 
\usepackage{siunitx}            
\usepackage{graphicx}           
\usepackage{xcolor}             
\usepackage{hyperref}           
\usepackage[version=3]{mhchem}  
\usepackage[mathlines]{lineno}  
\usepackage{epstopdf}           
\usepackage{calligra}           

\newcommand{\dd}{\,\text{d}}

\begin{document}

\title{Coherent and Incoherent Interfacial Spin Transport: Quantum-to-Classical Crossover in Spin Superfluids}

\author{A. R. Moura}
\email{antoniormoura@ufv.br}
\affiliation{Departamento de Física, Universidade Federal de Viçosa, 36570-900, Viçosa, Minas Gerais, Brazil}
\author{L. S. L. Barbosa}
\affiliation{Departamento de Física, Universidade Federal de Viçosa, 36570-900, Viçosa, Minas Gerais, Brazil}
\date{\today}

\begin{abstract}
We investigate the thermodynamics of interfacial spin transport within a normal metal/ferromagnetic insulator/normal metal ($\mathrm{NM/FMI/NM}$) trilayer heterostructure, where the central magnetic layer is described by the anisotropic quantum XXZ model. By employing the self-consistent harmonic approximation (SCHA) combined with a microscopic linear response formulation, we evaluate the interfacial spin-mixing conductance $g_{\uparrow\downarrow}$ across all spin regimes. We demonstrate that $g_{\uparrow\downarrow}$ uniquely decomposes into a coherent condensed component ($g_{\mathrm{cond}}$), driven by the macroscopic phase of the spin superfluid, and an incoherent fluctuation-driven term ($g_{\mathrm{fluct}}$) mediated by stochastic thermal magnons. Crucially, in the extreme quantum limit of $S = 1/2$, $g_{\mathrm{cond}}$ drops steeply and vanishes at a finite coherence temperature $T_{\mathrm{coh}}$. This singular quantum anomaly is triggered by severe local spin-flip fluctuations that disrupt phase coherence; it is regularized for higher spin values ($S \ge 1$) and completely quenched in the classical field limit ($S \to \infty$). Furthermore, we reveal that $T_{\mathrm{coh}}$ increases monotonically as the system transitions from the pure $\mathrm{XY}$ limit ($\lambda = 0$) toward the isotropic Heisenberg regime ($\lambda = 1$). This behavior originates from a geometric restructuring of the magnon modes from highly elliptic to circular precession, which hardens short-wavelength fluctuations at the zone boundary and expands the total magnon bandwidth, thereby shielding the macroscopic order parameter against thermal depletion. Conversely, the fluctuation-driven term $g_{\mathrm{fluct}}$ vanishes at $T = 0$, exhibits a characteristic $T^2$ quadratic scaling at low temperatures, and undergoes a systematic $1/S$ amplitude suppression as the macroscopic magnetization becomes robust. Our microscopic insights bridge the gap between quantum many-body fluctuations and macroscopic spin-superfluid hydrodynamics, providing clear foundational principles for optimizing long-range coherent transport in quantum spintronic devices.
\end{abstract}


\maketitle

\section{Introduction and motivation}
\label{sec:introduction}
Spin transport constitutes a cornerstone of modern spintronics research, with a particular emphasis on spin currents within magnetic insulators~\cite{ssp64.1}. In conventional architectures, spin-polarized charge currents are widely utilized to transport spin angular momentum within normal metal or semiconductor samples~\cite{rmp76.323}. However, intrinsic mechanisms such as Joule heating and rapid spin relaxation severely constrain the efficiency of conduction-electron spin transport, hindering its widespread implementation in nanoscale devices. Conversely, magnetic insulators have emerged as a promising alternative platform for mediating long-range spin transport. In these materials, spin propagation is mediated either by incoherent magnons (the quanta of spin waves)~\cite{prb65.172509,jpsj82.102002,natphys11.453} or by the coherent state of spin superfluids~\cite{advphys59.181,prb99.104423,ltp46.436,jmmm550.169033}. 

The classification of various physical phenomena as superfluid transport heavily depends on the precise definition of superfluidity~\cite{ltp46.436}. If a spatial phase gradient is considered the sole criterion for superfluid behavior, conventional acoustic waves, for instance, could be formally classified as a form of mass superfluid transport. In that case, however, the transport remains localized over short length scales comparable to the acoustic wavelength. To establish a physically rigorous framework, it is more convenient to adopt a definition of superfluidity that closely aligns with the foundational experiments in liquid $^4\mathrm{He}$~\cite{nature141.74,jnn12.2943}. Here, we define spin superfluidity as dissipationless transport sustained over macroscopic distances, where the phase gradient represents a necessary but not sufficient condition for achieving true superfluid currents.

To physically realize and detect this phenomenon in planar devices, a widely investigated architecture consists of a hybrid trilayer heterostructure, where an easy-plane ferromagnetic insulator (FMI) is sandwiched between two normal metal (NM) reservoirs acting as a spin source and a spin drain~\cite{prl112.227201}. In this open geometry, a spin current is injected at the left interface via spin-transfer torque (STT)~\cite{prb54.9353,jmmm159.l1}, driven by an electrically generated spin accumulation that establishes a non-uniform, spiral magnetic texture within the easy plane. The resulting spin current propagates along the longitudinal length $L_x$ of the FMI and is subsequently collected at the right interface through spin pumping (SP)~\cite{prl88.117601}. Within the framework of magnetoelectronic circuit theory~\cite{rmp34.694}, this process facilitates the quantification of the effective spin-mixing conductances across the heterostructure interfaces. Although the global magnetic precession within the easy plane is inherently subject to relaxation via Gilbert damping, the hydrodynamic description of spin superfluidity provides a robust analogy, provided that the spin-conservation laws are only weakly violated.

Despite the elegance of this hydrodynamic framework, the stability and thermodynamic behavior of spin superfluid transport at finite temperatures ($T > 0$) remain critical open questions. In realistic operating regimes, thermal fluctuations give rise to a substantial bath of incoherent magnons that strongly couple with the coherent spin condensate. Standard theoretical formulations typically treat these thermal contributions either within linear spin-wave theory or by employing continuous, long-wavelength effective field theories. However, such approximations fundamentally fail to capture the highly nonlinear, microscopic feedback that thermal excitations exert on the magnetic parameters. As the temperature increases, the nonlinear coupling among short-wavelength modes can significantly suppress the effective spin stiffness and strongly renormalize the local easy-plane anisotropy, potentially destabilizing the phase coherence required to sustain the macroscopic superfluid current.

To overcome the limitations of noninteracting spin-wave models and provide a more rigorous thermodynamic description of transport, a microscopic, nonperturbative approach is required. In this work, we employ the Self-Consistent Harmonic Approximation (SCHA) to systematically investigate the roles of both thermal and quantum fluctuations in spin superfluid transport. This variational scheme has a long history of successful application in low-dimensional magnetism, particularly in evaluating critical temperatures~\cite{prb49.9663,prb51.16413,prb59.6229}, Berezinskii-Kosterlitz-Thouless (BKT) transitions~\cite{prb48.12698,prb53.235,prb54.6081,prb78.212408}, and large-$D$ quantum phase transitions~\cite{jpcm20.015208,pasma388.3779}. Furthermore, the SCHA establishes a robust foundation for describing coherent states in FM systems and modern spintronic phenomena~\cite{jmmm472.1,prb106.054313,jmmm606.172393}. 

By representing the FMI via a discrete, anisotropic quantum spin Hamiltonian, the formalism maps the highly nonlinear thermal fluctuations onto an optimized, temperature-dependent effective harmonic Hamiltonian. Notably, recent developments in this method have extended its capabilities to rigorously treat quantum spin models, providing a powerful and efficient tool for investigating the thermodynamics of $S = 1/2$ systems~\cite{jmmm655.174331}. This optimized variational framework accounts for the self-consistent renormalization of both the microscopic exchange couplings and the local anisotropy fields, thereby capturing the nonlinear feedback of the thermal bath far beyond the constraints of conventional linear spin-wave or effective long-wavelength theories. Utilizing this framework, we systematically evaluate how the thermal bath influences spin current propagation, the local phase profile, and the effective spin-mixing conductances across the heterostructure interfaces.

This paper is organized as follows. In Sec.~\ref{sec:model}, we define the microscopic Hamiltonian of the NM/FMI/NM heterostructure and outline the self-consistent variational equations of the SCHA. In Sec.~\ref{sec:sfbg}, we establish the fundamental features of superfluid transport from a semiclassical perspective, while Sec.~\ref{sec:magcircuit} details the magnetoelectronic circuit modeling of spin currents across the heterostructure. In Sec.~\ref{sec:scthermo}, we derive expressions for the thermally renormalized spin-mixing conductances and discuss the interplay between the spin condensate and thermal fluctuations. Section~\ref{sec:results} presents our numerical results and a detailed analysis of the stability limits of the superfluid state under thermal effects. Finally, our concluding remarks and prospective applications for spintronic devices are summarized in Sec.~\ref{sec:conclusions}.

\section{Model description}
\label{sec:model}
In this work, we investigate the thermodynamics of spin transport through a trilayer structure comprising a FMI sandwiched between two NM reservoirs. The NM/FMI interface is modeled as a rectangular region with an area of $A_\perp = L_y L_z$ lying in the $yz$ plane, while the FMI layer extends along the $x$ axis with a longitudinal length of $L_x \gg a$, where $a$ denotes the lattice constant of the FM sample. The left NM is maintained in a non-equilibrium state characterized by a spin-dependent chemical potential imbalance between spin-up and spin-down electrons, $\Delta\mu = \mu_\uparrow - \mu_\downarrow > 0$. 

\begin{figure}[ht]
\centering
\includegraphics[width=0.9\linewidth]{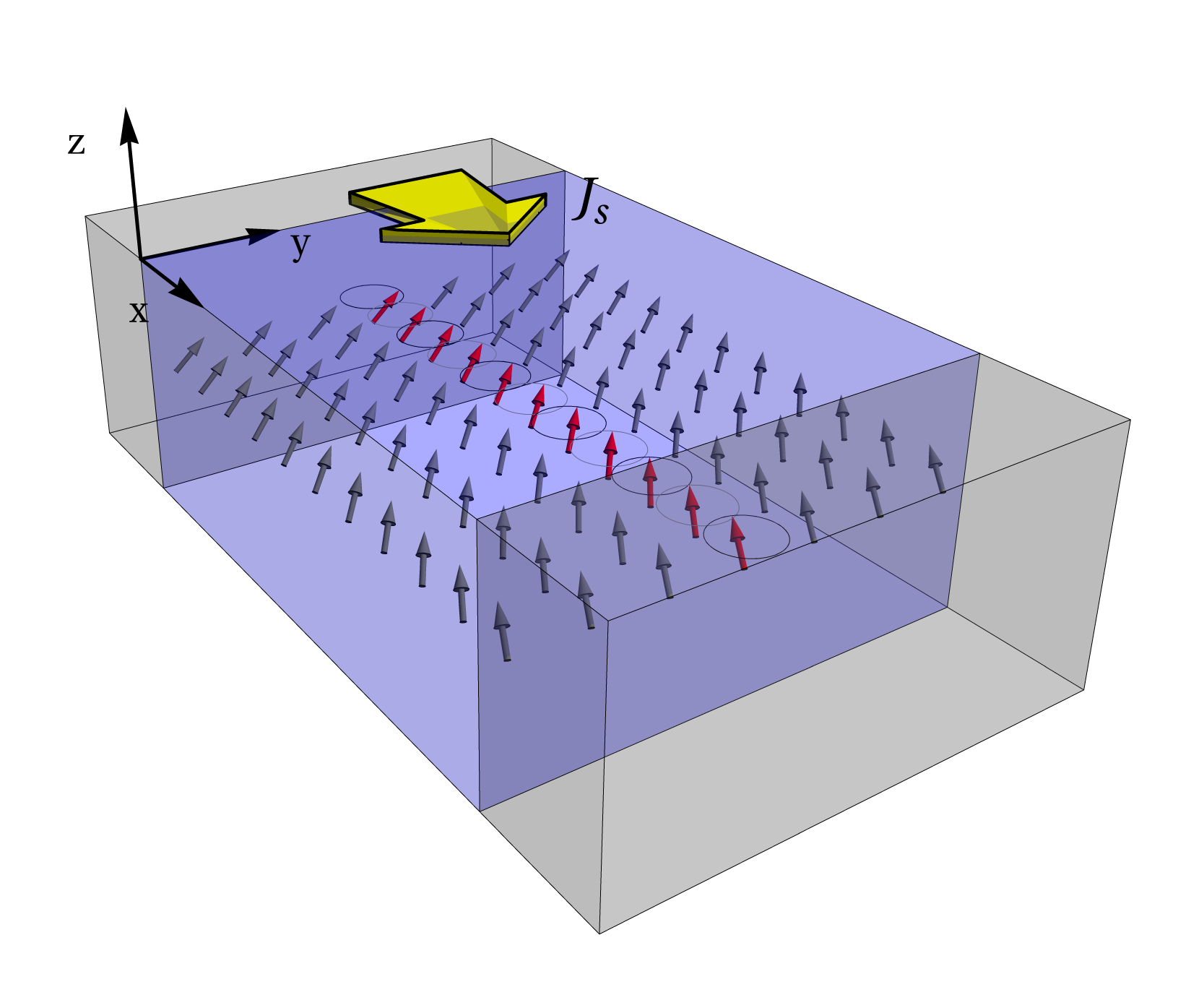}
\caption{Schematic illustration of a spin-superfluid state in an easy-plane FMI (blue region) sandwiched between two NM reservoirs (gray regions). The coherent transport of angular momentum is mediated by a spatial variation of the magnetic phase, resulting in a spin supercurrent $J_s$ flowing along the $x$ direction. The red arrows depict the local magnetization, while the gray circles indicate its precessional texture within the easy plane. A spin current density $J_s$ is injected from the left NM into the FMI via STT.}
\label{fig:sample}
\end{figure}

In this scenario, the spin current crossing the NM/FMI interface is intrinsically a non-equilibrium effect arising from the finite chemical potential imbalance between the spin-up and spin-down electronic populations in the left NM reservoir. In contrast, thermodynamic equilibrium imposes the condition $\mu_\uparrow = \mu_\downarrow$, which precludes the existence of any net spin current. Such a spin-dependent chemical potential imbalance can be maintained, for instance, by spin accumulation generated through spin-orbit coupling. At the microscopic level, within this biased regime, spin is injected from the NM into the FMI via scattering processes that involve spin-flip electronic reflections accompanied by magnon emission or absorption.

The Hamiltonian model considered in this work includes contributions from the NM reservoirs, the FMI, and the interfacial interaction. The Hamiltonian associated with the left NM reservoir is described by the free-electron model:
\begin{align}
    \label{eq:He}
    H_{e,L} = \sum_{k\sigma} \xi_{k\sigma} c_{k\sigma}^\dagger c_{k\sigma},
\end{align}
where the single-particle excitation energy is given by $\xi_{k\sigma} = \epsilon_{k} - \mu_{\sigma}$, and $c_{k\sigma}^{\dagger}$ ($c_{k\sigma}$) denotes the fermionic creation (annihilation) operator for an electron state with momentum $k$ and spin projection $\sigma = \uparrow, \downarrow$. Since the right metallic reservoir is treated as an ideal spin sink, the Hamiltonian $H_{e,R}$ is defined analogously; however, it does not include any spin-dependent chemical potential imbalance induced by external sources.

The local $s$-$d$ exchange interaction couples the conduction electron spin operators of the NM layer, denoted by $\mathbf{s}_{\ell}$, with the localized spin operators $\mathbf{S}_{\ell}$ of the FMI. Although this interfacial exchange coupling formally includes both longitudinal and transverse components, only the transverse channels contribute to the transfer of angular momentum. Accordingly, the interaction Hamiltonian can be expressed as:
\begin{align}
    H_{sd} = -\frac{J_{sd}}{2} \sum_{\ell \in \text{int}} \left( S_\ell^+ s_\ell^- + S_\ell^- s_\ell^+\right),
\end{align}
where $J_{sd}$ characterizes the magnitude of the $s$-$d$ exchange interaction, and the summation extends over all lattice sites on the interface ($yz$ plane) indexed by $\ell$. The electronic spin-raising operator is expressed in terms of the fermionic creation and annihilation operators as $s_\ell^+ = c_{\ell\uparrow}^{\dagger} c_{\ell\downarrow}$.

The FMI is modeled by the XXZ Hamiltonian, expressed as follows:
\begin{align}
    \label{eq:Hm}
    H_m = -J \sum_{\langle ij \rangle} \Biggl( \frac{S_i^+ S_j^- + S_i^- S_j^+}{2} + \lambda S_i^z S_j^z \Biggr),
\end{align}
where $0 \le \lambda < 1$ denotes the easy-plane anisotropy that energetically favors spin configurations confined to the $xy$ plane, and the sum runs over nearest-neighbor pairs. Given that an isotropic model yields a quadratic energy spectrum, which, according to the Landau criterion, precludes the sustained support of a superfluid spin current, the easy-plane configuration becomes indispensable. Additionally, in the long-wavelength limit, the magnon spectrum is given by
\begin{align}
    \hbar\omega_q \approx 2JSaq \sqrt{z(1-\lambda) + \lambda a^2 q^2}.
\end{align}
The crossover between the low-energy linear regime ($\omega_q \propto q$) and the high-energy quadratic regime ($\omega_q \propto q^2$) takes place when the two energy scales inside the square root become comparable, namely $\lambda a^2 q^2 \sim z(1-\lambda)$. This condition defines a characteristic crossover temperature
\begin{align}
    T_\lambda \sim \frac{JS}{k_B}\frac{1-\lambda}{\sqrt{\lambda}}.
\end{align}

For weak anisotropy ($\lambda \approx 1$), $T_\lambda$ is much smaller than the Curie temperature $T_C \sim JS^2$, yielding a broad temperature window in which thermally excited magnons behave as nearly circular modes and recover the quadratic dispersion characteristic of an isotropic ferromagnet. In contrast, in the XY limit ($\lambda \approx 0$), the crossover scale diverges and moves beyond the magnon bandwidth, meaning that the spectrum remains effectively linear throughout the physically accessible energy range. Consequently, the notion of a crossover from linear to quadratic magnons is meaningful only sufficiently close to the Heisenberg point $\lambda = 1$, whereas for strongly anisotropic easy-plane magnets, the Goldstone-like linear dispersion dominates the magnon dynamics at all relevant temperatures. Accordingly, $T_\lambda$ should not be interpreted as a phase-transition temperature, but rather as the energy scale above which the effects of easy-plane anisotropy become negligible in magnon dynamics. In the present study, we focus on the regime defined by strong anisotropy.

Motivated by the semiclassical approach in terms of the canonically conjugate fields $\varphi$ and $S^z$, which enable a transparent description of spin hydrodynamics, we adopt the Villain representation~\cite{jp35.27} to express the spin operator as $S_i^+ = e^{i\varphi_i}\sqrt{\tilde{S}^2 - S_i^z(S_i^z - 1)}$, where $\tilde{S}^2 = S(S - 1)$. The fields $\varphi$ and $S^z$ are dimensionless operators that satisfy the canonical commutation relation $[\varphi_i, S_j^z] = i\delta_{ij}$. For a smoothly varying spin field characterized by $\langle\Delta\varphi\rangle \ll 1$, along with a small out-of-plane spin component, it is justified to expand Eq.~\eqref{eq:Hm} as a quadratic model in the $\varphi$ and $S^z$ operators. Furthermore, to incorporate spin-wave interactions that are neglected within the harmonic approximation, we introduce a renormalization factor $\rho$ that accounts for both thermal and quantum fluctuations~\cite{jmmm655.174331}. 

Assuming that the longitudinal length $L_x$ is kept finite while the transverse area $L_y L_z$ is taken to the thermodynamic limit $L_y L_z \to \infty$, the field operator can be expanded as follows:
\begin{align}
    \varphi_i = \int_{\!-\pi/a}^{\pi/a} \! \dd q_y \int_{\!-\pi/a}^{\pi/a} \! \dd q_z \sum_{q_x=\pi/L_x}^{\pi/a} \varphi_q \,\psi_q(\mathbf{r}_i).
\end{align}
In this representation, the transverse momentum components $q_y$ and $q_z$ are treated as continuous variables within the first Brillouin zone, whereas the longitudinal component $q_x$ is discretized according to $q_x = \pi m/L_x$, $m = 1, \dots, N_x - 1$. This discretization reflects the finite system size $L_x = (N_x - 1)a$ along the $x$ direction. In the above expression, we introduce a plane-wave-like basis function
\begin{align}
    \psi_q(\mathbf{r}_i) = \frac{1}{\sqrt{\mathcal{N}}} e^{i \mathbf{q}_\perp \cdot \mathbf{r}_\perp} c_{q_x} \cos(q_x x_i),
\end{align}
where $\mathcal{N} = (2\pi/a)^2(N_x - 1)$ and we employ the discrete cosine transform of type I (DCT-I) to account for the finite extent and open boundary conditions along the $x$ direction. In this framework, the coefficients $c_{q_x}$ serve as normalization factors that reflect the boundary geometry: at the boundary modes $q_x = 0$ and $q_x = \pi/a$, one has $c_{q_x} = 1$, whereas for the interior modes within the interval $0 < q_x < \pi/a$, they take the value $c_{q_x} = \sqrt{2}$. 

By employing this complete basis set and performing the standard diagonalization procedure, the FMI Hamiltonian in momentum space is expressed in the quadratic form as
\begin{equation}
H_m^0 = \frac{1}{2} \sum_q \bigl( h_q^\varphi \overline{\varphi}_q \varphi_q + h_q^z \overline{S}_q^z S_q^z \bigr),
\end{equation}
where $h_q^\varphi=2zJ\tilde{S}^2\rho(1-\gamma_q)$ and $h_q^z=2zJ(1-\lambda\gamma_q)$. For notational convenience, we adopt the convention that the summation over momentum is implicitly understood as an integration over the transverse momentum components, combined with an ordinary discrete summation over the longitudinal component. The lattice geometry is encoded in the structure factor $\gamma_q = z^{-1}\sum_\eta e^{i\mathbf{q}\cdot\boldsymbol{\eta}}$, which accounts for the $z$ nearest-neighbor spins located at positions $\boldsymbol{\eta}$. The renormalization parameter is determined from the self-consistent equation~\cite{jmmm655.174331}
\begin{align}
    \rho(T)=\Lambda(T)\left(1-\frac{\langle (S^z)^2\rangle_0}{\tilde{S}^2} \right)e^{-\tfrac{1}{2}\langle\Delta\varphi^2\rangle_0},
\end{align}
where the function
\begin{align}
    \Lambda(T)=\Biggl[\sum_q\Biggl(v_q-\frac{1-e^{-\tfrac{1}{2}\langle\Delta\varphi^2\rangle_0}}{1-\lambda\gamma_q}u_q \Biggr) \Biggr]^{-1}\sum_qv_q
\end{align}
represents a strictly quantum correction, while $u_q$ and $v_q$ are functions of the magnon energy $\hbar\omega_q$ given by
\begin{align}
    \label{eq.u}
    u_q=\hbar\omega_q\frac{\sinh(\beta\hbar\omega_q)-\beta\hbar\omega_q}{[2\sinh(\beta\hbar\omega_q/2)]^2},
\end{align}
and
\begin{align}
    \label{eq.v}
    v_q=\hbar\omega_q\frac{\sinh(\beta\hbar\omega_q)+\beta\hbar\omega_q}{[2\sinh(\beta\hbar\omega_q/2)]^2},
\end{align}
with $\beta=(k_B T)^{-1}$. The thermal averages $\langle\ldots\rangle_0$ are computed with respect to the non-interacting Hamiltonian $H_m^0$. In the semiclassical limit, appropriate for large spin quantum values, $\Lambda(T)$ approaches unity, and the expectation values are governed by Gaussian distributions.

\section{Superfluid background state}
\label{sec:sfbg}
To analyze the thermodynamic properties of the superfluid spin current, it is convenient to formulate the spin-field dynamics by taking the classical solution as a reference configuration. Subsequently, quantum fluctuations are incorporated as small perturbations around this classical background.

To determine the classical solution, it is advantageous to represent the Hamiltonian in the continuum limit. Since our interest primarily lies in the low-temperature regime, we operate in the long-wavelength limit, for which the harmonic Hamiltonian in real space is expressed as:
\begin{equation}
    H_m^\mathrm{cl} = \frac{1}{2}\int \bigl[ A (\boldsymbol{\nabla} \varphi)^2 + K (n^z)^2 \bigr] \, \dd^3r,
\end{equation}
where $A = 2J\tilde{S}^2\rho/a$ is the renormalized spin stiffness and $K = 2zJ\tilde{S}^2 (1-\lambda)/a^3$ parametrizes the easy-plane anisotropy. In these expressions, we assume that $\varphi$ and $n^z = S^z/\tilde{S}$ are classical ($c$-number) fields. Additionally, since $n^z \ll 1$, we disregard all terms containing the gradient $\boldsymbol{\nabla} n^z$. The corresponding dynamical equations can therefore be written in the following form:
\begin{equation}
    {\calligra s}\dot{\varphi} = K n^z + \alpha {\calligra s}\dot{n}^z,
\end{equation}
and
\begin{equation}
    {\calligra s}\dot{n}^z = A\nabla^2 \varphi - \alpha {\calligra s}\dot{\varphi},
\end{equation}
where ${\calligra s}=\hbar\tilde{S}/a^3$ denotes the spin density. Following the Landau-Lifshitz-Gilbert (LLG) formalism, we have included a phenomenological dissipative contribution characterized by the dimensionless Gilbert damping parameter $\alpha \ll 1$. The solution of this coupled system of equations yields the linear dispersion relation $\omega_q = (1 - i\alpha)\, q c$, where the spin-wave velocity is given by $c = 2J\tilde{S}a\sqrt{z\rho(1-\lambda)}/\hbar$.

In the steady-state regime, and in the absence of spin-superfluidity contributions, the transport of non-equilibrium thermal magnons within the FMI bulk ($0 \le x \le L_x$) is governed by the one-dimensional spin diffusion equation, $\nabla^2\mu_m = \mu_m/\lambda_m^2$, where $\mu_m(x)$ denotes the magnon spin chemical potential and $\lambda_m$ is the bulk magnon spin diffusion length~\cite{duine,prb94.014412}. Note that to define $\mu_m(x)$, it is assumed that the characteristic scattering time associated with magnon-magnon interactions is significantly shorter than the corresponding timescale for magnon-lattice scattering processes. Additionally, the spatial profile of $\mu_m(x)$ is uniquely determined by the magnetoelectronic boundary conditions at the NM/FMI interfaces. At the left interface ($x = 0$), the diffusive magnon spin current density, defined by $\mathbf{J}_m = -\sigma_m \boldsymbol{\nabla}\mu_m$, must satisfy the self-consistent boundary condition $J_m(0) = g_\mathrm{hyb} \left[\Delta\mu_\mathrm{oc} - \mu_m(0)\right]/4\pi$. Here, $\sigma_m$ is the bulk magnon spin conductivity, $\Delta\mu_\mathrm{oc}$ characterizes the open-circuit spin accumulation driven by the adjacent NM reservoir, and $g_{\mathrm{hyb}}$ is the effective hybrid conductance that encapsulates both the NM bulk relaxation and the interfacial spin-mixing conductance (further details are provided in Sec.~\ref{sec:magcircuit}). Analogously, the right interface ($x = L_x$) imposes the boundary condition $J_m(L_x) = g_\mathrm{hyb}\,\mu_m(L_x)/4\pi$. Solving this boundary value problem yields the transmitted spin current density that reaches the right FMI/NM interface:
\begin{align}
    \label{eq:Jm}
    J_m(L_x) = \frac{g_\mathrm{hyb}^2 g_m \,\Delta\mu_\mathrm{oc}}{4\pi}\Biggr[2 g_m g_{\mathrm{hyb}} \cosh\!\biggl(\frac{L_x}{\lambda_m}\biggr)\nonumber\\
        +(g_m^2+ g_{\mathrm{hyb}}^2) \sinh\!\biggl(\frac{L_x}{\lambda_m}\biggr)\Biggr]^{-1},
\end{align}
where $g_m = 4\pi\sigma_m/\lambda_m$ denotes the intrinsic spin conductance of the bulk magnon transport channel. For YIG at room temperature, experimental measurements yield $\sigma_m \approx 10^{10}\,\mathrm{m}^{-1}$ and $\lambda_m \approx 9.4\times10^{-6}\,\mathrm{m}$, which correspond to a conductance of $g_m \approx 1.4\times10^{16}\,\mathrm{m}^{-2}$~\cite{natphys11.1022}. In particular, when this length is derived from the microscopic XXZ spin model, the magnon diffusion length takes the explicit form $\lambda_m = (a/\alpha)\sqrt{\lambda/[z(1-\lambda)]}$, which predicts length scales on the order of $10^{-5}\ \mathrm{m}$. As anticipated, the spin diffusion signal exhibits pronounced attenuation as the thickness of the FMI layer increases.

Within this hydrodynamic framework, the macroscopically induced precessional frequency $\Omega$ plays a profound role analogous to a dynamic spin chemical potential. Just as an electrical voltage or a conventional chemical potential gradient drives mass or charge transport, the uniform precession of the planar magnetization at frequency $\Omega$ acts as the fundamental driving force that sustains the spin-superfluid transport. This precessional frequency establishes a Josephson-like relation, where the time-evolution of the condensate phase is locked to the out-of-plane spin accumulation, enabling long-range coherent spin transfer across the bulk channel without relying on the population of thermal quasiparticles. Alternatively, to obtain long-range superfluid spin transport, we consider a steady-flux configuration that is independent of the transverse spatial coordinates and search for classical solutions satisfying $\dot{n}_\mathrm{cl}^z(x,t)=0$ and $\varphi_\mathrm{cl}(x,t)=\phi(x)+\Omega t$~\cite{prl112.227201}. Consequently, we obtain the macroscopically induced precession frequency $\Omega=Kn^z/{\calligra s}$ and the spin continuity equation $\boldsymbol{\nabla}\cdot\mathbf{J}_s=-\alpha{\calligra s}\Omega$, where $\mathbf{J}_s=-A\boldsymbol{\nabla}\varphi_\mathrm{cl}$ represents the superfluid spin current density vector. By applying the divergence theorem along the longitudinal direction, we find $\Delta J_s = J_s^\text{L} - J_s^\text{R} = \alpha{\calligra s}\Omega L_x$, where $J_s$ denotes the scalar longitudinal component of the current density. The required boundary conditions are determined by the dynamical spin state at the interfaces, as detailed in Sec.~\ref{sec:magcircuit}.

\section{Self-Consistent Boundary Spin Currents}
\label{sec:magcircuit}
When the left NM reservoir is driven into a non-equilibrium state, a spin-transfer torque (STT)~\cite{prb54.9353,jmmm159.l1} is exerted at the interface, thereby injecting a spin current into the FMI. For maximal efficiency, we assume that the injected spin current $I_\mathrm{in}^\mathrm{L} = J_\mathrm{in}^\mathrm{L}A_\perp$ carries angular momentum polarized along the $z$ axis. Consequently, the spin current density injected from the left is given by:
\begin{align}
J_\mathrm{in}^\mathrm{L}=\frac{1}{4\pi}\bigl(\mathrm{Im}\,g_{\uparrow\downarrow}^\mathrm{L}+\mathrm{Re}\,g_{\uparrow\downarrow}^\mathrm{L}\,\mathbf{n}\times\bigr)\bigl(\Delta\boldsymbol{\mu}_\mathrm{L}\times\mathbf{n}\bigr),
\end{align}
where $\mathbf{n}$ denotes the classical magnetization unit vector obtained from the semiclassical reference configuration, $g_{\uparrow\downarrow}^\mathrm{L}$ is the left interfacial spin-mixing conductance, and $\Delta\boldsymbol{\mu}_\mathrm{L}=\Delta\mu_\mathrm{L}\,\hat{\mathbf{z}}$ represents the spin-dependent chemical potential difference in the left reservoir. Once the local magnetization near the interface is tilted by the injected spin flux, the spins undergo macroscopically coherent precession. This magnetization dynamics gives rise to spin pumping (SP)~\cite{prl88.117601}, which constitutes the exact Onsager-reciprocal process of STT. As a result, a finite fraction of the injected spin angular momentum is pumped back into the left NM reservoir. The corresponding back-flowing spin current density is expressed as
\begin{align}
J_\mathrm{out}^\text{L}=\frac{\hbar}{4\pi}\bigl(\mathrm{Im}\,g_{\uparrow\downarrow}^\mathrm{L}+\mathrm{Re}\,g_{\uparrow\downarrow}^\mathrm{L}\,\mathbf{n}\times\bigr)\bigl[(\mathbf{n}\times\dot{\mathbf{n}})\times\mathbf{n}\bigr].
\end{align}
The spin-mixing conductance can be evaluated within the scattering-matrix formalism and, for typical metallic interfaces, satisfies $\mathrm{Im}\,g_{\uparrow\downarrow}^\mathrm{L} \ll \mathrm{Re}\,g_{\uparrow\downarrow}^\mathrm{L}$~\cite{rmp77.1375}. The net spin current density injected across the left boundary then simplifies to $J_s^\mathrm{L}=g_{\uparrow\downarrow}^\mathrm{L}\bigl(\Delta\mu_\mathrm{L}-\hbar\Omega\bigr)/4\pi$. An analogous boundary analysis at $x = L_x$ yields the net transmitted spin current density $J_s^\mathrm{R} = g_{\uparrow\downarrow}^\mathrm{R}(\hbar \Omega-\Delta\mu_\mathrm{R}) / 4\pi$. For simplicity, we consider symmetric interfaces and set $g_{\uparrow\downarrow}^\mathrm{L} = g_{\uparrow\downarrow}^\mathrm{R} = g_{\uparrow\downarrow}$. 

To establish a direct link between this theoretical framework and experimentally accessible transport signatures, we model the spin accumulation within the adjacent NM reservoirs by explicitly incorporating the reciprocal spin Hall mechanisms. In this layout, a longitudinal charge current density $\mathbf{J}_c$ traversing the left injector reservoir generates a transverse spin current via the Spin Hall Effect (SHE)~\cite{prl83.1834}, characterized by $\mathbf{J}_\mathrm{SH} = \theta_{\mathrm{SH}} (2e/\hbar)(\boldsymbol{\sigma}\times \mathbf{J}_c)$, where $\theta_{\mathrm{SH}}$ represents the spin Hall angle and $\boldsymbol{\sigma}$ denotes the spin polarization unit vector. Reciprocally, within the right NM detector reservoir, the Inverse Spin Hall Effect (ISHE) converts the transmitted spin flux back into a measurable transverse charge current density~\cite{jap97.10c715,apl88.182509}.

For a finite NM layer localized within the region $-d \leq x \leq 0$, the spin accumulation is parametrized by the spin chemical potential $\mu_s(x) = \Delta\mu(x)/2$. The corresponding diffusive spin current is given by $\mathbf{J}_\text{diff}(x) = -(\hbar \sigma_e / 2e^2) \boldsymbol{\nabla}\mu_s(x)$, where $\sigma_e$ is the bulk electrical conductivity~\cite{rezende}. The total spin current density, $J_s(x) = J_\mathrm{SH} + J_\text{diff}(x)$, satisfies the steady-state spin diffusion equation $\nabla^2 \mu_s(x) = \mu_s(x)/\lambda_\text{sf}^2$, where $\lambda_\mathrm{sf}$ denotes the bulk spin diffusion length. This boundary-value problem is subject to the open-boundary condition $J_s(-d) = 0$ at the external surface and continuity at the interface, $J_s(0) = J_s^\mathrm{L}$. Solving this system yields the self-consistent spin chemical potential imbalance at the injector boundary:
\begin{align}
    \Delta\mu_\mathrm{L}=\frac{4\pi}{g_\mathrm{NM}}\bigl(J_\mathrm{SH}^\mathrm{net}-J_s^\mathrm{L}\bigr),
\end{align}
where $J_\mathrm{SH}^\mathrm{net} =\bigl[1-\mathrm{sech}(d/\lambda_\mathrm{sf}) \bigr] J_\mathrm{SH}$ represents the net driven spin current that survives bulk spin-flip relaxation to reach the interface, and
\begin{align}
    \label{eq:gNM}
    g_\mathrm{NM} = \frac{\pi \hbar \sigma_e}{e^2 \lambda_\mathrm{sf}} \tanh\!\biggl(\frac{d}{\lambda_\mathrm{sf}}\biggr)    
\end{align}
denotes the intrinsic spin conductance of the finite NM layer. By substituting $\Delta\mu_\mathrm{L}$ back into the expression for the interfacial spin current, we find
\begin{align}
    J_s^\mathrm{L}=\frac{g_\mathrm{hyb}}{4\pi}\bigl(\Delta\mu_{\mathrm{oc}}-\hbar\Omega\bigr),
\end{align}
where $g_\mathrm{hyb}=g_{\uparrow\downarrow}g_\mathrm{NM}/(g_{\uparrow\downarrow}+g_\mathrm{NM})$ denotes the hybrid spin-mixing conductance of the combined interface-reservoir channel. The quantity $\Delta\mu_{\mathrm{oc}}=(4\pi/g_\mathrm{NM}) J_\mathrm{SH}^\mathrm{net}$ corresponds to the open-circuit spin accumulation, representing the maximum achievable spin accumulation in the limit of a fully blocked interface.

In the right NM reservoir ($L_x \leq x \leq L_x + d$), where no direct spin Hall drive is active, the spin diffusion equation is subject to the boundary conditions $J_s(L_x) = J_s^\mathrm{R}$ and $J_s(L_x + d) = 0$. This results in a non-equilibrium accumulation $\Delta\mu_\mathrm{R} = (4\pi/g_\mathrm{NM}) J_s^\mathrm{R}$. Accounting for the finite transparency of the right interface, the profile of the spin current pumped into the right NM detector layer reads:
\begin{align}
    \label{eq:JsRx}
    J_s(x)=J_s^\mathrm{R}\frac{\sinh\bigl[(L_x+d-x)/\lambda_\mathrm{sf}\bigr]}{\sinh\bigl[d/\lambda_\mathrm{sf}\bigr]},
\end{align}
with $J_s^\mathrm{R}=(g_\mathrm{hyb}/4\pi)\hbar\Omega$. Finally, the conservation of angular momentum within the FMI bulk dictates that the spatial discontinuity between the boundary spin currents must contribute to the internal magnetic relaxation, such that $J_s^\mathrm{L}-J_s^\mathrm{R}=g_\alpha\hbar\Omega/4\pi$, where $g_\alpha=4\pi\alpha {\calligra s} L_x/\hbar$ is the bulk damping conductance. By combining these boundary conditions, we solve the coupled system to determine the exact amplitude of the spin current density transmitted to the right NM reservoir, given by $J_s^\mathrm{R}=g_\mathrm{hyb}^2\Delta\mu_{\mathrm{oc}}/[4\pi(2g_\mathrm{hyb}+g_\alpha)]$. Crucially, in contrast to the diffusive spin current described by Eq.~\eqref{eq:Jm}, which decays exponentially, the superfluid spin current exhibits robust algebraic decay along the FMI channel. By invoking Onsager reciprocity and spatially averaging the pumped spin current profile of Eq.~\eqref{eq:JsRx} across the normal metal detector layer, we obtain the measurable non-local charge current ratio:
\begin{align}
    \label{eq:Jcratio}
    \biggl|\frac{J_c^\mathrm{R}}{J_c^\mathrm{L}}\biggr|=&\frac{G_0\theta_\mathrm{SH}^2\lambda_\mathrm{sf}^2}{2\sigma_e d}\tanh^2\biggl(\frac{d}{2\lambda_\mathrm{sf}}\biggr)\frac{g_\mathrm{hyb}^2}{2g_\mathrm{hyb}+g_\alpha},
\end{align}
where $G_0 = e^2/h$ denotes the fundamental quantum of electrical conductance.

The foregoing result is derived in the zero-temperature limit, whereas finite-temperature effects give rise to additional contributions. It is important to note that our analysis explicitly incorporates the renormalization parameter $\rho$, which plays a crucial role in determining the thermodynamic properties of $g_{\uparrow\downarrow}$, as will be examined in detail in the following section. 

Notably, in the asymptotic limit where the intrinsic spin conductance of the NM layer far exceeds the interfacial transparency, $g_\mathrm{NM} \gg g_{\uparrow\downarrow}$, the hybrid spin-mixing conductance reduces to $g_\mathrm{hyb} \approx g_{\uparrow\downarrow}$. In this regime, the diffusive spin-accumulation drop within the bulk reservoir becomes negligible, and our self-consistent framework asymptotically recovers the conventional results established by Takei and Tserkovnyak~\cite{prl112.227201}, where the reservoir is treated under the idealized spin-sink approximation.

\section{Spin current thermodynamics}
\label{sec:scthermo}
Once the classical solution has been determined, we can incorporate quantum fluctuations. Accordingly, we adopt the previously obtained classical solution as the reference configuration with respect to which spin fluctuations are defined and analyzed. We then decompose the conjugate operators as $\varphi(\mathbf{r}) = \varphi_\mathrm{cl}(\mathbf{r}) + \delta\varphi(\mathbf{r})$ and $S^z(\mathbf{r}) = \tilde{S} n_\mathrm{cl}^z(\mathbf{r}) + \delta S^z(\mathbf{r})$, treating the classical contributions as $c$-number background fields, while the deviations represent quantum operators that satisfy the canonical commutation relation $[\delta\varphi(\mathbf{r}), \delta S^z(\mathbf{r}^\prime)] = i \delta^3(\mathbf{r}-\mathbf{r}^\prime)$. Under this decomposition, the unperturbed magnetic Hamiltonian is given by $H_m^0 = E_\mathrm{cl} + (1/2) \sum_q \bigl( h_q^\varphi \, \delta\bar{\varphi}_q \delta\varphi_q + h_q^z \, \delta\bar{S}_q^z \delta S_q^z \bigr)$, where $E_\mathrm{cl}$ denotes the energy associated with the classical background configuration. The quantum Hamiltonian is diagonalized by defining bosonic operators as follows:
\begin{subequations}
\label{eq:aadagger}
\begin{align}
    \delta\varphi_q&=\frac{1}{\sqrt{2}}\Biggl(\frac{h_q^z}{h_q^\varphi} \Biggr)^{1/4}(a_q^\dagger+a_{-q})\\
    \delta S_q^z&=\frac{i}{\sqrt{2}}\Biggl(\frac{h_q^\varphi}{h_q^z} \Biggr)^{1/4}(a_q^\dagger-a_{-q}),
\end{align}    
\end{subequations}
resulting in the harmonic model
\begin{align}
    H_m^0=E_\mathrm{cl}+\sum_q \hbar\omega_q\Biggl(a_q^\dagger a_q+\frac{1}{2}\Biggr),
\end{align}
where $\hbar\omega_q=2zJ\tilde{S}\sqrt{\rho(1-\gamma_q)(1-\lambda\gamma_q)}$. In the long-wavelength limit, we approximate the structure factor as $\gamma_q\approx 1 - a^2q^2/z$, and the magnon energy assumes the linear form $\hbar\omega_q=\hbar qc$, where $c$ denotes the spin-wave velocity determined in Sec.~\ref{sec:sfbg}.

Based on the superfluid background, the spin-raising operator can be approximated as $S_i^+\approx\tilde{S}e^{i(\varphi_\mathrm{cl}+\delta\varphi_i)}$, such that the transverse spin-spin correlation function $D_{ij}^{-+}(t)=\langle S_i^-(t)S_j^+(0)\rangle_0$ takes the form
\begin{align}
    \label{eq:Dij}
    D_{ij}^{-+}(t)\approx \tilde{S}^2 e^{i\Delta\phi} \exp\!\Bigl[-\tfrac{1}{2}\langle\bigl(\delta\varphi_i(t)-\delta\varphi_j(0)\bigr)^2\rangle_0\Bigr],
\end{align}
where $\Delta\phi=\phi_j-\phi_i$. To derive this expression, we employed the quadratic approximation of the magnetic Hamiltonian, which allows us to replace the average over phase fluctuations with a Gaussian distribution. It should be noted that the thermodynamic average is expressed in terms of the deviation operators $\delta\varphi$ and $\delta S^z$, which together give rise to the harmonic form of the Hamiltonian.

For small phase deviations, we expand the correlation function given by Eq.~\eqref{eq:Dij} up to second order in $\delta\varphi$. In terms of the bosonic operators, the non-local phase correlation function reads:
\begin{align}
    \langle\delta\varphi_i(t)\delta\varphi_j(0)\rangle_0=&\,\frac{\hbar}{4zJ\tilde{S}^2\rho}\,\frac{1}{\mathcal{N}} \sum_q \frac{\omega_q e^{i\mathbf{q}\cdot\Delta\mathbf{r}}}{1-\gamma_q}\Bigl[n_q e^{i\omega_q t}\nonumber\\
    &+ (n_q+1)e^{-i\omega_q t}\Bigr],
\end{align}
where $\Delta\mathbf{r}=\mathbf{r}_j-\mathbf{r}_i$, and $n_q=n(\hbar\omega_q)$ denotes the Bose-Einstein distribution for magnons with energy $\hbar\omega_q$. Consequently, the spin-spin correlation function in frequency-momentum space can be approximated as:
\begin{align}
    \label{eq:Dtildeij}
    \tilde{D}_{ij}^{-+}(\varepsilon)=&\,2\pi\hbar\tilde{S}^2\delta(\varepsilon)-\frac{\pi\hbar^2}{2zJ\rho\,\mathcal{N}}\sum_q e^{i\mathbf{q}\cdot\Delta\mathbf{r}}\frac{\omega_q}{1-\gamma_q}\nonumber\\
    &\times\big[(2n_q+1)\delta(\varepsilon)-n_q\delta(\varepsilon+\hbar\omega_q)\nonumber\\
    &-(n_q+1)\delta(\varepsilon-\hbar\omega_q)\bigr],
\end{align}
where $\varepsilon=\hbar\nu$.

For sufficiently small chemical potential imbalances, the $s$-$d$ exchange interaction can be treated as a perturbation over an equilibrium state, and the resulting spin current injected across the interface can be described using linear-response theory. In this formalism, $\Delta \mu$ acts as the generalized thermodynamic force that drives the spin current across the left interface ($x=0$). Since spin-flip scattering at the interface leads to a decrease in the population of spin-up electrons and a concomitant increase in the population of spin-down electrons, we define the spin current operator as $I_s=(\hbar/2)\sum_k\left(\dot{n}_{k\downarrow}-\dot{n}_{k\uparrow}\right)$, where $n_{k\sigma} = c_{k\sigma}^\dagger c_{k\sigma} $ denotes the electron number operator. Because the electron number operator commutes with both the FMI and NM Hamiltonians, the time evolution of $n_{k\sigma}$ is governed solely by the commutator $[n_{k\sigma},H_{sd}]$. The Heisenberg equation of motion thus yields a compact expression for the injected spin current:
\begin{align}
    I_s=\frac{2 i J_{sd}}{N_e}\sum_{\ell k p}\bigl(A_{\ell kp}^\dagger-A_{\ell kp} \bigr),
\end{align}
where $N_e$ is the number of electrons in the reservoir, and we define the mixed operator $A_{\ell kp}=S_\ell^-c_{k\uparrow}^\dagger c_{p\downarrow} e^{-i(\mathbf{k}-\mathbf{p})\cdot\mathbf{r}_\ell}$. Note that $\mathbf{r}_\ell$ designates a spatial position restricted to the $yz$ plane, corresponding to the boundary condition $x=0$.

In the interaction picture, the expectation value of the spin current is given by $\langle I_s(t)\rangle = \langle \mathcal{S}^\dagger(t)\,\hat{I}_s(t)\,\mathcal{S}(t)\rangle_0$, where the hat indicates time evolution with respect to the non-interacting Hamiltonian, and $\mathcal{S}(t) = T_t \exp\left[-\frac{i}{\hbar}\int_{-\infty}^t \hat{H}_{sd}(t')\,\dd t'\right]$ denotes the $\mathcal{S}$ matrix, with $T_t$ representing the time-ordering operator. For weak interfacial exchange coupling, the $\mathcal{S}$ matrix can be perturbatively expanded to first order in $H_{sd}$, yielding the injected spin current density:
\begin{align}
    J_s^\mathrm{L}=\frac{g_{\uparrow\downarrow}^\mathrm{L}}{4\pi}(\Delta\mu-\hbar\Omega),
\end{align}
where the spin-mixing conductance is defined as
\begin{align}
    \label{eq:GL}
    g_{\uparrow\downarrow}^\mathrm{L} = -\frac{8\pi J_{sd}^2}{\Delta\mu-\hbar\Omega}\frac{1}{A_\perp}\sum_{\ell\ell^\prime}\frac{1}{N_e^2}\sum_{kp}\mathrm{Im}\,\tilde{\chi}_{\ell\ell^\prime kp}(0).
\end{align}
In the above expression, $\tilde{\chi}_{\ell\ell^\prime kp}(\omega)$ is the temporal Fourier transform of the retarded Green's function defined by $\chi_{\ell\ell^\prime kp}(t)=(i\hbar)^{-1}\theta(t)\big\langle [A_{\ell kp}^\dagger(t),A_{\ell^\prime kp}(0)]\big\rangle_0$. In Appendix~\ref{appendix}, we demonstrate that after carrying out the momentum and position summations, we obtain
\begin{align}
    \label{eq:imchitilde}
    \mathrm{Im}\,\tilde{\chi}(0)=&-\frac{1}{2\hbar}\bigl(1-e^{-\beta(\Delta\mu-\hbar\Omega)}\bigr)\,\sum_\ell\tilde{C}_{\ell\ell}(0),
\end{align}
where $\tilde{C}_{\ell\ell^\prime}(\omega)$ defines the temporal Fourier transform of the dynamical correlation function
\begin{align}
    C_{\ell\ell^\prime}(t)=\frac{1}{N_e^2}\sum_{kp}\big\langle\hat{A}_{\ell kp}(t)\,\hat{A}_{\ell^\prime kp}^\dagger(0)\big\rangle_0.
\end{align}
Note that Eq.~\eqref{eq:imchitilde} is formally analogous to the fluctuation-dissipation theorem. However, the assumption of detailed balance is not applied, as the system under consideration does not reside in thermodynamic equilibrium.

The determination of the correlation function $C_{\ell\ell^\prime}(t)$ plays a central role in evaluating the spin current. By employing the definition of the mixed operator, we find that
\begin{align}
    C_{\ell\ell^\prime}(t) =&\frac{1}{N_e^2}\sum_{kp} \langle c_{k\uparrow}^\dagger c_{p\downarrow} c_{p\uparrow} c_{p\downarrow}^\dagger \hat{S}_\ell^-(t)\hat{S}_{\ell^\prime}^+(0)\rangle_0\nonumber\\
    &\times e^{i(\mathbf{p}-\mathbf{k})\cdot\Delta\mathbf{r}} e^{i(\nu_k-\nu_p)t},
\end{align}
where $\Delta \mathbf{r}=\mathbf{r}_\ell-\mathbf{r}_{\ell^\prime}$ denotes a displacement vector restricted to the $yz$ plane, and $\nu_{(k,p)} = \epsilon_{(k,p)}/\hbar$. It is convenient to replace the discrete sum over $k$ and $p$ with integrals over the full momentum space. Furthermore, since the spin and electronic degrees of freedom are decoupled, we employ Wick's theorem to obtain
\begin{align}
    \tilde{C}_{\ell\ell^\prime}(0)=&\frac{V_e^2}{N_e^2}\int\frac{\dd^3k}{(2\pi)^3}\int\frac{\dd^3p}{(2\pi)^3}\,f_{k\uparrow}(1-f_{p\downarrow})e^{i(\mathbf{p}-\mathbf{k})\cdot\Delta\mathbf{r}}\nonumber\\
    &\times\int\dd\varepsilon \tilde{D}_{\ell\ell^\prime}^{-+}(\varepsilon)\delta(\varepsilon-\epsilon_k+\epsilon_p),
\end{align}
where $f_{k\sigma}=f(\xi_{k\sigma})=(e^{\beta\xi_{k\sigma}}+1)^{-1}$ denotes the Fermi-Dirac distribution and $V_e$ is the NM reservoir volume. Here, $\varepsilon$ is associated with the magnon energy, and therefore the Dirac delta function ensures energy conservation during the corresponding scattering processes. Under the constraint of energy conservation, we employ the identity
\begin{align}
    f_{k\uparrow}(1-f_{p\downarrow})=\frac{f_{p\downarrow}-f_{k\uparrow}}{e^{\beta(\varepsilon-\Delta\mu)}-1}
\end{align}
to simplify the expression. The angular integrations can be evaluated straightforwardly by noting that
\begin{align}
    \int_0^{2\pi}\dd\vartheta\int_0^\pi\dd\theta \sin\theta\, e^{-ik\Delta r\cos\theta}=4\pi\,\mathrm{sinc}(k\Delta r),
\end{align}
where $\mathrm{sinc}(x)=x^{-1}\sin x$ holds an analogous result for the $p$ integral. For a typical NM, the conduction electrons are tightly constrained near the Fermi surface ($k \approx k_F \sim 10^{10}\,\mathrm{m}^{-1}$), meaning that the factor $\mathrm{sinc}(k_F \Delta r)$ becomes appreciable only in the local limit $\Delta r \to 0$. In particular, by taking $\Delta r$ to be on the order of the lattice spacing $a$, one obtains $\mathrm{sinc}(k_F a) \sim 10^{-2}$, which fully justifies the local approximation $\mathrm{sinc}^2(k_F \Delta r) \approx \delta_{\ell\ell'}$. 

Next, the momentum integrations over $k$ and $p$ are transformed into energy integrals over the respective electronic density of states. One of these energy integrals can be simplified due to the presence of the Dirac delta function. Furthermore, in the low-temperature regime, we employ the approximation $f(\xi - \Delta\xi) - f(\xi) \approx \Delta\xi\,\delta(\xi)$ (valid for $\Delta\xi\ll\xi$), which allows us to eliminate the second energy integral through an additional delta function constraint. Consequently, assuming a constant absolute density of states at the Fermi level $N(\epsilon_F)=V_e mk_F/\pi^2\hbar^2=3N_e/2\epsilon_F$, we obtain:
\begin{align}
    \label{eq:sumCll}
    \sum_\ell\tilde{C}_{\ell\ell}(0)=\frac{N(\epsilon_F)^2N_\perp}{4N_e^2} \int \tilde{D}^{-+}(\varepsilon) \frac{\varepsilon-\Delta\mu}{e^{\beta(\varepsilon-\Delta\mu)}-1}\dd\varepsilon,
\end{align}
where $\tilde{D}^{-+}(\varepsilon)=N_\perp^{-1}\sum_\ell\tilde{D}_{\ell\ell}^{-+}(\varepsilon)$ represents the spatially averaged transverse spin correlation function at the interface. By substituting Eqs.~\eqref{eq:GL}, \eqref{eq:imchitilde}, and \eqref{eq:sumCll} into this relation, we obtain
\begin{align}
    \label{eq:gLud}
    g_{\uparrow\downarrow}^\text{L}(T)=&\frac{\pi}{\hbar}\Biggl(\frac{J_{sd}N(\epsilon_F)}{aN_e}\Biggr)^2 \Theta(T)\int\tilde{D}^{-+}(\varepsilon)\nonumber\\
    &\times(\varepsilon-\Delta\mu)\, n(\varepsilon-\Delta\mu)\dd\varepsilon,
\end{align}
where we define $\Theta(T)=(1-e^{-\beta(\Delta\mu-\hbar\Omega)})/(\Delta\mu-\hbar\Omega)$. 

To determine the spin current injected at the right interface ($x = L_x$), where the spin chemical potential bias vanishes, one can exploit the spatial boundary symmetries and global spin conservation instead of recalculating the entire linear response scheme. Physically, while the left interface acts as a spin source, the right interface functions as a spin sink. To ensure steady-state spin transport through the NM/FMI/NM system, the right NM must experience a positive rate of spin accumulation, meaning that $\dot{N}_{\uparrow} > 0$ at $x=L_x$. This inversion in the net directional flow of angular momentum manifests in the frequency domain as an effective sign reversal of the driving energies, mapping $\varepsilon \to -\varepsilon$ (or equivalently, $\omega \to -\omega$). Furthermore, since the macroscopically induced precession frequency within the easy-plane ferromagnet scales directly with the local magnetization along the quantization axis, $\Omega \propto S^z$, the relative orientation of the spin transfer torque at the right boundary requires replacing $\Omega$ with $-\Omega$. This joint transformation effectively mirrors the boundary condition results previously obtained for the left interface without violating causality or the dissipative nature of the transport equations.

Finally, upon substituting the expression for the spin-spin correlation function given in Eq.~\eqref{eq:Dtildeij}, the spin-mixing conductance can be decomposed into two distinct contributions as:
\begin{align}
    g_{\uparrow\downarrow}(T) = g_{\mathrm{cond}}(T) + g_{\mathrm{fluct}}(T),
\end{align}
which holds for both interfaces. The first term is given by
\begin{align}
    g_{\mathrm{cond}}(T) = g_{\uparrow\downarrow}^0 \Biggl( 1-\frac{F_1(\lambda)}{2\tilde{S}\sqrt{\rho(T)}} \Biggr),
\end{align}
and denotes the condensate contribution, as it remains finite even in the limit of zero temperature. Here, $g_{\uparrow\downarrow}^0=2(\pi J_{sd}\tilde{S}N(\epsilon_F)/aN_e)^2$ represents the spin-mixing parameter at zero temperature, evaluated in the classical limit $S\to\infty$. Using typical values of the material parameters, namely $\epsilon_F \sim 5\,\mathrm{eV}$, $J_{sd}\sim 0.1\,\mathrm{eV}$, and a lattice constant $a \sim 10^{-10}\,\mathrm{m}$, we estimate $g_{\uparrow\downarrow}^0$ to be on the order of $10^{18}\,\mathrm{m}^{-2}$, which is in agreement with the reported experimental values~\cite{prl107.066604,jap111.07c307,pra1.044004}. The function $F_1(\lambda)$ is a temperature-independent quantity defined by
\begin{align}
    F_1(\lambda) = \frac{1}{\mathcal{N}}\sum_q \sqrt{\frac{1 - \lambda \gamma_q}{1 - \gamma_q}}.
\end{align}
For small values of the anisotropy parameter, $F_1(\lambda)$ can be expanded in a power series in $\lambda$, as detailed in Appendix~\ref{appendix2}. The second term characterizes the contribution resulting from magnon fluctuations and is given by
\begin{align}
    g_{\mathrm{fluct}}(T)= g_{\uparrow\downarrow}^0\frac{F_2(\lambda,\Delta\mu,T)}{2zA(T)\Delta\mu},
\end{align}
where $A(T)=2J\tilde{S}^2\rho(T)/a$, and we define the function
\begin{align}
    F_2(\lambda,\Delta\mu,T)=&\frac{1}{\mathcal{N}} \sum_{q}\frac{\hbar\omega_q}{1-\gamma_q}\Biggl(\frac{\hbar\omega_q-\Delta\mu}{e^{\beta(\hbar\omega_q-\Delta\mu)}-1}\nonumber\\
    &- \frac{\hbar\omega_q+\Delta\mu}{e^{\beta(\hbar\omega_q+\Delta\mu)}-1}\Biggr).
\end{align}
It should be noted that, in contrast to the contribution emerging from the condensed component, the fluctuation spin-mixing conductance vanishes in the zero-temperature limit. Additionally, observe that $F_2$ also vanishes in the classical limit, which is obtained by keeping $A$ fixed while taking $S \to \infty$. Consequently, $g_\mathrm{fluct}$ captures thermal corrections that are of order $\tilde{S}^{-1}$. Owing to the exponential suppression of high-energy magnon modes at low temperatures, the magnon dispersion can be accurately approximated by its linear low-energy spectrum, which yields the following analytical expression:
\begin{align}
    g_{\text{fluct}}(T)=&g_{\uparrow\downarrow}^0\frac{(k_B T)^2}{(2\pi)^2 A(T) \hbar c}\Biggl[\frac{\pi^2}{6}-\biggl(\frac{\epsilon_D}{k_B T}\biggr)^2 \text{Li}_0(e^{-\beta\epsilon_D})\nonumber\\
    &-\biggl(\frac{\epsilon_D}{k_B T}\biggr)\text{Li}_1(e^{-\beta\epsilon_D})-\text{Li}_2(e^{-\beta\epsilon_D}) \Biggr],
\end{align}
where $\epsilon_D=(6\pi^2)^{1/3}\hbar c/a$ represents the ultraviolet cutoff energy, and $\mathrm{Li}_s$ denotes polylogarithmic functions of order $s$. A detailed derivation of the above expressions is provided in Appendix~\ref{appendix2}.

\section{Results}
\label{sec:results}
\begin{figure}[ht]
\centering
\includegraphics[width=\linewidth]{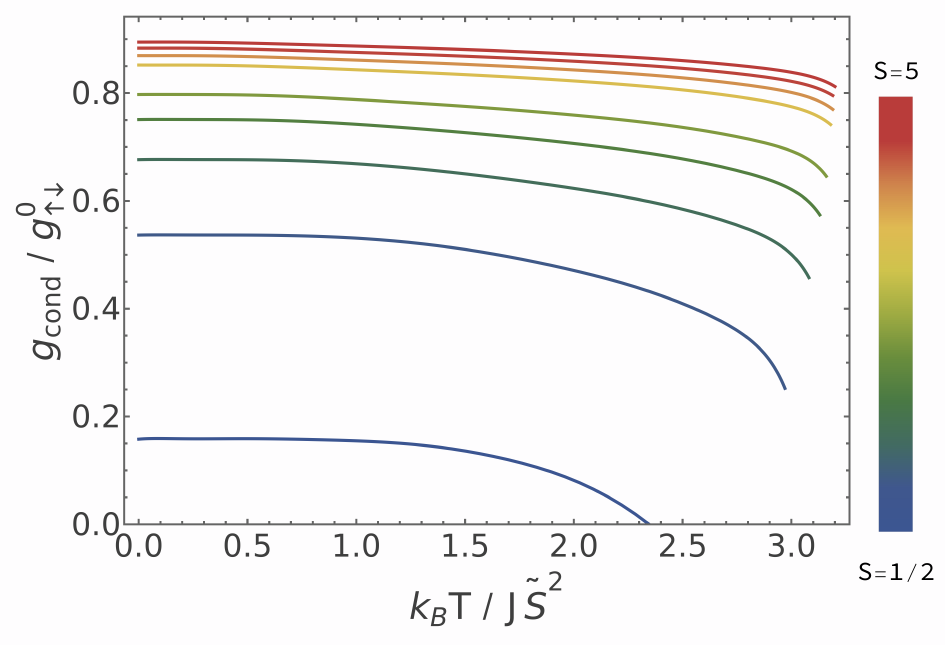}
\caption{Temperature dependence of the coherent condensed spin-mixing conductance $g_{\mathrm{cond}}$ for multiple spin values in the range $S = 1/2$ to $S = 5$ (in increments of $\Delta S = 1/2$) with anisotropy $\lambda=0.1$. In particular, the $S = 1/2$ case exhibits a complete vanishing of the coherent conductance at a finite coherence temperature $k_B T_{\mathrm{coh}} \approx 2.3 J\tilde{S}^2$, where $\tilde{S}$ is the effective spin magnitude.}
\label{fig:gcond}
\end{figure}

The condensed component $g_{\mathrm{cond}}$, which characterizes the coherent spin transport mediated directly by the macroscopic phase of the spin superfluid, is plotted in Fig.~\ref{fig:gcond} as a function of temperature for various spin quantum numbers $S$. At $T = 0$, $g_{\mathrm{cond}}$ reaches its maximum value, representing an ideal coherent channel across the interface. As the temperature increases, $g_{\mathrm{cond}}$ undergoes monotonic thermal depletion due to the gradual destruction of planar magnetic order by thermal fluctuations. 

An interesting quantum anomaly emerges in the extreme quantum limit of $S = 1/2$, where the coherent conductance $g_{\mathrm{cond}}$ drops steeply and vanishes completely at a finite coherence temperature $T_{\mathrm{coh}} \approx 0.85 T_c$. Notably, this boundary phase coherence collapses before the complete thermal degradation of the renormalization parameter $\rho$. Microscopically, a single spin-flip scattering event mediated by the interfacial $J_{sd}$ coupling represents a total reversal of the local magnetization, resulting in a $100\%$ inversion for a spin-1/2 localized moment. Such severe fluctuations disrupt the phase coherence of the adjacent spin condensate. Consequently, within the critical temperature window $T_{\mathrm{coh}} \le T \le T_c$, the interface effectively suppresses the superfluid channel, $g_{\mathrm{cond}} \to 0$, forcing the non-local spin transport to be sustained exclusively via the incoherent magnon diffusion channel $g_{\mathrm{fluct}}$.

\begin{figure}[ht]
\centering
\includegraphics[width=0.9\linewidth]{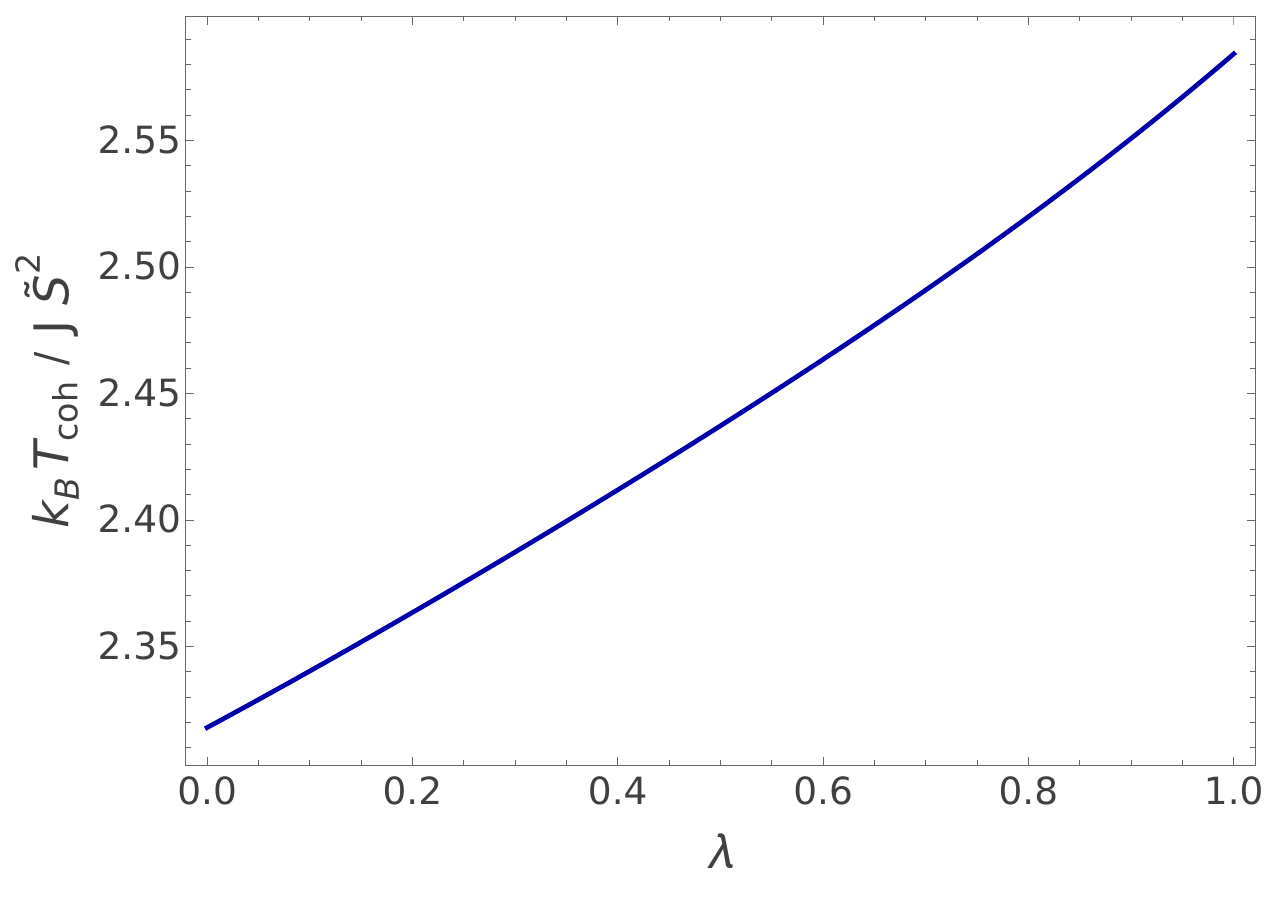}
\caption{Anisotropy dependence of the coherence temperature $T_{\mathrm{coh}}$ for the extreme quantum limit $S = 1/2$, highlighting the threshold above which the condensed spin-mixing conductance $g_{\mathrm{cond}}$ vanishes.}
\label{fig:Tcoh}
\end{figure}

Additionally, as shown in Fig.~\ref{fig:Tcoh}, this coherence temperature exhibits a weak but monotonic increase as the system transitions from the pure XY limit ($\lambda = 0$) to the isotropic Heisenberg regime ($\lambda = 1$). To provide a deeper physical insight into this trend, we analyze the geometric and energetic restructuring of the magnon spectrum driven by $\lambda$. In the XY-dominated regime, out-of-plane fluctuations are highly penalized, forcing the excitations to become highly elliptic and soft at the zone boundary, with a maximum energy bounded by $\hbar\omega_{\mathrm{max}} = 2\sqrt{2\rho}zJ\tilde{S}$, where $\rho$ denotes the spin-stiffness reduction factor. Conversely, as $\lambda \to 1$, the restoration of spatial isotropy allows the magnons to become circular, effectively hardening the short-wavelength fluctuations and expanding the total magnon bandwidth up to $\hbar\omega_{\mathrm{max}} = 4zJ\sqrt{\rho}\tilde{S}$. Consequently, at higher values of $\lambda$, the high-energy magnon modes require a larger thermal activation energy to be populated. This quenches the thermal depletion of the macroscopic order parameter, thereby shifting the quantum collapse of the coherent channel to a higher coherence temperature $T_{\mathrm{coh}}$.

For higher spin values ($S \ge 1$), the background macroscopic magnetization becomes increasingly robust against individual electronic scattering events, thereby shielding the coherent transport channel. Consequently, $g_{\mathrm{cond}}$ remains finite over a broader temperature range. In the classical limit ($S \gg 1$), the condensed conductance becomes entirely immune to thermal depletion, asymptotically converging to the temperature-independent constant $g_{\uparrow\downarrow}^0$. This flat behavior corroborates the classical field-theoretical description of spin superfluids, wherein quantum shot noise at the boundary is completely suppressed.

In contrast to the condensed component, the fluctuation-driven term $g_{\mathrm{fluct}}$ captures the incoherent transport channels mediated by stochastic thermal excitations at the interface. At $T = 0$, where the ferromagnet is frozen into its quantum ground state, the absence of thermal magnons dictates that $g_{\mathrm{fluct}} = 0$. For $T > 0$, $g_{\mathrm{fluct}}$ grows monotonically with temperature, reflecting the progressive thermal population of non-equilibrium spin-flip channels. In the regime of very low temperatures, we find that $g_{\mathrm{fluct}}$ scales as $T^2$, in excellent agreement with the behavior previously reported by Takei and Tserkovnyak in Ref.~\onlinecite{prl112.227201}.

\begin{figure}[ht]
\centering
\includegraphics[width=0.95\linewidth]{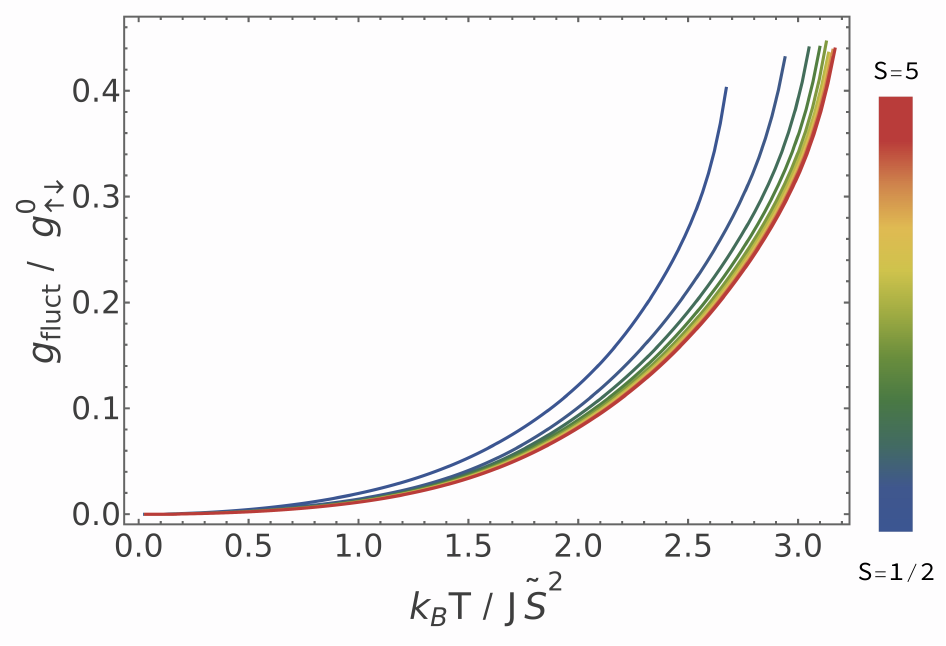}
\caption{Contribution arising from the fluctuation-induced spin-mixing conductance $g_{\mathrm{fluct}}$ as a function of temperature for various spin values. In the low-temperature regime, the curves exhibit a characteristic quadratic scaling, $g_{\mathrm{fluct}} \propto T^2$.}
\label{fig:gfluct}
\end{figure}

Our results reveal a systematic ordering of the fluctuation curves as a function of the spin quantum number $S$. The amplitude of $g_{\mathrm{fluct}}$ is maximum for $S = 1/2$ and undergoes steady suppression for larger spin values. This behavior characterizes $g_{\mathrm{fluct}}$ as an intrinsically quantum-mechanical correction. As the system approaches the semiclassical regime, the localized spins increasingly behave as rigid classical vectors, thereby quenching the discrete fluctuation channels. In the strict classical limit, the fluctuation component vanishes entirely, ensuring that the total conductance $g$ smoothly matches the classical macrospin prediction $g_{\uparrow\downarrow}^0$ at all temperatures.

To verify our results within a realistic experimental layout, we analyze a heavy-metal reservoir composed of $\mathrm{Pt}$, for which the typical spin diffusion length and electrical conductivity are $\lambda_{\mathrm{sf}} \approx 1.5\,\mathrm{nm}$ and $\sigma_e \approx 5\times 10^6\,(\Omega\mathrm{m})^{-1}$, respectively~\cite{prb87.224401,prl116.126601}. Under these conditions, the intrinsic spin conductance of the bulk metal reaches $g_{\mathrm{NM}} \sim 10^{19}\,\mathrm{m}^{-2}$ for thicknesses $d \gtrsim \lambda_{\mathrm{sf}}$. Given that the interfacial spin-mixing conductance $g_{\uparrow\downarrow}$ is typically restricted to the order of $10^{18}\,\mathrm{m}^{-2}$, we can safely apply the ideal spin-sink approximation, setting $g_{\mathrm{hyb}} \approx g_{\uparrow\downarrow}$. Consequently, the absolute ratio of the converted charge currents can be expressed in the following compact scaling form:
\begin{align}
    \Biggl|\frac{J_c^{\mathrm{R}}}{J_c^{\mathrm{L}}}\Biggr|_\mathrm{sf} = \eta(d/\lambda_\mathrm{sf}) \frac{G_0\theta_{\mathrm{SH}}^2\lambda_{\mathrm{sf}}}{\sigma_e}\frac{g_{\uparrow\downarrow}^2}{2g_{\uparrow\downarrow}+g_\alpha},
\end{align}
where $G_0 = e^2/h$ denotes the quantum of conductance, $\eta(r)=(2/r)\tanh^2(r/2)$ defines the dimensionless geometric efficiency factor, and $g_\alpha$ represents the effective bulk conductance associated with Gilbert damping losses. This function encapsulates the competitive trade-off between efficient interfacial spin absorption and bulk electrical shunting, attaining its global maximum value of $\eta_{\mathrm{max}} \approx 0.582$ at $r \approx 2.177$ (corresponding to an optimal operational thickness of $d \approx 2.2\,\lambda_{\mathrm{sf}}$). Conversely, assuming purely diffusive spin transport, the resulting expression for the ratio of charge currents is given by
\begin{align}
\Biggl|\frac{J_c^{\text{R}}}{J_c^{\text{L}}}\Biggr|_\text{diff} =&\,\eta(d/\lambda_\text{sf}) \frac{G_0\theta_{\text{SH}}^2\lambda_{\text{sf}}}{\sigma_e}g_\text{m}g_{\uparrow\downarrow}\bigl[g_{\uparrow\downarrow}\sinh(L_x/\lambda_\text{m})\nonumber\\
&+2g_\text{m}\cosh(L_x/\lambda_\text{m})]^{-1},
\end{align}
where we take $g_m \ll g_{\uparrow\downarrow}$.

\begin{figure}[ht]
\centering
\includegraphics[width=0.95\linewidth]{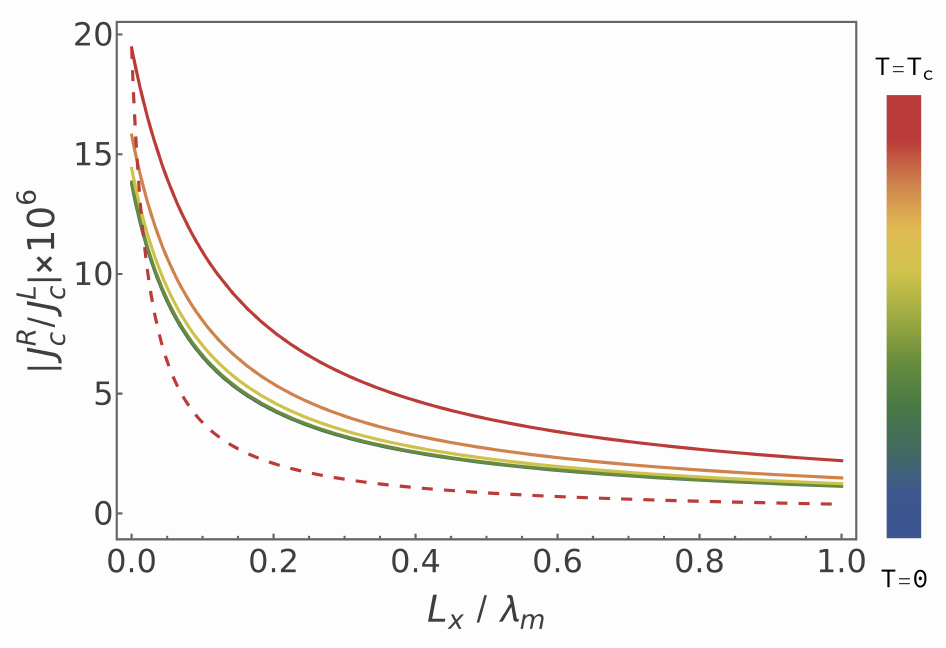}
\caption{Non-local charge current transmission ratio $|J_c^{\mathrm{R}}/J_c^{\mathrm{L}}|$ as a function of the normalized FM length $L_x / \lambda_m$. The dashed lines track the theoretical prediction for incoherent diffusive magnon transport, while the solid lines represent the coherent spin-superfluid regime. The vertical color bar maps the temperature evolution of the transport channels from the absolute zero limit ($T = 0$, blue) up to the critical temperature ($T = T_c$, red). The spatial scale is normalized by the room-temperature magnon diffusion length, $\lambda_m = 9.4\ \mu\mathrm{m}$.}
\label{fig:Jratio}
\end{figure}

Fig.~\ref{fig:Jratio} illustrates the non-local resistance efficiency, quantified by the charge current ratio $|J_c^{\mathrm{R}}/J_c^{\mathrm{L}}|$, as a function of the normalized ferromagnetic channel length $L_x / \lambda_m$ across the entire temperature landscape up to $T_c$. A striking physical contrast emerges between the two transport mechanisms. The conventional diffusive magnon contribution (represented by the dashed lines) suffers from severe exponential attenuation governed by $\exp(-L_x/\lambda_m)$. Crucially, as the system approaches the ultra-low temperature regime ($T \to 0$), this diffusive channel undergoes a complete thermal freeze-out, vanishing entirely due to the suppression of the magnon spin conductance ($g_m \to 0$). In stark contrast, spin-superfluid transport (depicted by the solid lines) bypasses this spatial limitation, exhibiting a highly robust algebraic decay. Remarkably, at the absolute zero limit ($T = 0$), where conventional magnon diffusion is fully deactivated, spin superfluidity remains operational and delivers a finite, long-range spin signal. This persistence underscores that the collective precessional transport is mediated by the macroscopically coherent ground-state condensate rather than thermal fluctuations. Furthermore, as the temperature is swept towards $T_c$ (indicated by the thermal color gradient), the superfluid efficiency undergoes systematic modulation driven by the temperature dependence of the microscopic XXZ parameters, yet it consistently outperforms the diffusive baseline over large distance scales.

\section{Summary and Conclusions}
\label{sec:conclusions}
In summary, we have developed a comprehensive microscopic framework to investigate the thermodynamics of interfacial spin transport within an anisotropic NM/FMI/NM trilayer heterostructure. By implementing the SCHA framework, we successfully mapped the highly nonlinear thermal fluctuations of the XXZ ferromagnetic insulator onto an optimized, temperature-dependent effective Hamiltonian. This variational approach allowed us to evaluate the interfacial spin current density via a microscopic linear response formulation without resorting to premature bosonic mappings, thereby fully preserving the quantum kinematic constraints of the localized spin operators across all spin-quantum-number regimes.

A central achievement of this work is the exact microscopic derivation and decomposition of the total spin-mixing conductance into two physically distinct components: a coherent condensed contribution ($g_{\mathrm{cond}}$) and an incoherent fluctuation-driven term ($g_{\mathrm{fluct}}$). The condensed conductance $g_{\mathrm{cond}}$, which dictates the transport mediated by the macroscopic phase of the spin superfluid, undergoes a monotonic thermal depletion from its absolute maximum at $T = 0$. Crucially, in the extreme quantum limit of $S = 1/2$, $g_{\mathrm{cond}}$ drops steeply and vanishes identically at a finite coherence temperature $T_{\mathrm{coh}}$. We demonstrated that this complete collapse of the coherent channel is a unique quantum anomaly triggered by severe local spin-flip fluctuations, where a single scattering event causes a $100\%$ local magnetization inversion, coupled with the rigid algebra of spin-1/2 operators. For higher spins ($S \ge 1$), this anomaly is systematically regularized by the robust field-theoretical background of the larger macrospin, eventually converging to a flat, temperature-independent constant $g_{\uparrow\downarrow}^0$ in the strict classical limit ($S \to \infty$).

Furthermore, by embedding these quantum interfacial conductances into a self-consistent non-local transport geometry, we established the macroscopic device efficiency through the charge current transmission ratio $|J_c^{\mathrm{R}}/J_c^{\mathrm{L}}|$ as a function of the normalized channel length $L_x / \lambda_m$. This global analysis revealed a striking physical dichotomy between the two competing transport regimes. While the conventional incoherent magnon channel suffers from a severe exponential attenuation penalty $\sim \exp(-L_x/\lambda_m)$ and undergoes complete thermal freeze-out ($g_m \to 0$) as $T \to 0$, the spin-superfluid transport effectively bypasses these thermal and geometric constraints. The precessional transport exhibits a highly robust algebraic decay, remaining fully operational down to the absolute zero limit. The persistence of a finite, long-range spin signal at $T = 0$ provides an unambiguous experimental signature of a macroscopically coherent ground-state condensate rather than thermally activated fluctuations. Ultimately, our findings provide a rigorous theoretical roadmap for identifying and optimizing spin superfluidity in low-dimensional quantum magnetic insulators, establishing a clear pathway toward long-range, dissipationless quantum spintronic signaling.

This study was financed in part by the Coordenação de Aperfeiçoamento de Pessoal de Nível Superior (CAPES) - Finance Code 001, and by the Conselho Nacional de Desenvolvimento Científico e Tecnológico (CNPq). 

\appendix
\section{Derivation of Interfacial Susceptibility} 
\label{appendix}
The imaginary part of the temporal Fourier transform of the retarded Green's function,
\begin{equation}
    \tilde{\chi}_{\ell\ell^\prime kp}(\omega)=\int_{-\infty}^{\infty} \chi_{\ell\ell^\prime kp}(t)e^{i\omega t}\,\dd t,
\end{equation}
can be expressed in terms of its time-domain components as
\begin{align}
    \operatorname{Im}\tilde{\chi}_{\ell\ell^\prime kp}(\omega)=\frac{1}{2i}\int_{-\infty}^{\infty} \bigl[\chi_{\ell\ell^\prime kp}(t)-\bar{\chi}_{\ell\ell^\prime kp}(-t)\bigr]e^{i\omega t}\,\dd t.
\end{align}
Performing the summation over the electronic momentum channels $k$ and $p$, we obtain the spatially resolved imaginary susceptibility:
\begin{align}
    \operatorname{Im}\tilde{\chi}_{\ell\ell^\prime}(\omega)=\frac{\delta_{\ell\ell^\prime}}{2\hbar}\bigl[\tilde{C}_{\ell\ell^\prime}^{+-}(\omega)-\tilde{C}_{\ell\ell^\prime}^{-+}(\omega)\bigr],
\end{align}
where we define the core correlation functions as $C_{\ell\ell^\prime}(t)=C_{\ell\ell^\prime}^{-+}(t)=N_e^{-2}\sum_{kp}\bigl\langle\hat{A}_{\ell kp}(t)\,\hat{A}_{\ell^\prime kp}^\dagger(0)\bigr\rangle_0$ and $C_{\ell\ell^\prime}^{+-}(t)=N_e^{-2}\sum_{kp}\bigl\langle\hat{A}_{\ell kp}^\dagger(t)\,\hat{A}_{\ell^\prime kp}(0)\bigr\rangle_0$. Furthermore, we have employed the local interface condition $C_{\ell\ell^\prime}(t)\approx \delta_{\ell\ell^\prime} C_{\ell\ell}(t)$, which holds true for both correlation functions due to the localized nature of the $s$-$d$ exchange coupling. 

To establish the exact relationship between the spectral functions $\tilde{C}_{\ell\ell}^{-+}({\omega})$ and $\tilde{C}_{\ell\ell}^{+-}(\omega)$, we examine the boundary condition imposed by the thermal bath in the grand canonical ensemble:
\begin{align}
    &\bigl\langle \hat{A}_{\ell kp}^\dagger(0)\hat{A}_{\ell kp}(t)\bigr\rangle_0 =\frac{1}{Z_0}\operatorname{Tr}\left[e^{-\beta K_0} \hat{A}_{\ell kp}^\dagger(0)\hat{A}_{\ell kp}(t)\right] \nonumber \\
    &=e^{-\beta(\Delta\mu-\hbar\Omega)}\bigl\langle \hat{A}_{\ell kp}(t-i\beta\hbar)\hat{A}_{\ell kp}^\dagger(0)\bigr\rangle_0,
\end{align}
where $K_0=H_e-\sum_{k\sigma} \mu_\sigma c_{k\sigma}^\dagger c_{k\sigma}+H_m^0$ denotes the unperturbed grand canonical Hamiltonian of the heterostructure. Provided that $[H_e+H_m^0,\sum_\sigma\mu_\sigma N_\sigma]=0$, we invoke the identity $e^{\beta K_0}\hat{A}_{\ell kp}(t)e^{-\beta K_0}=e^{-\beta\Delta\mu}\hat{A}_{\ell kp}(t-i\beta\hbar)$ to account for the electronic chemical potential shift. Finally, by applying the temporal Fourier transform to this boundary relation, we derive the generalized fluctuation-dissipation scaling:
\begin{equation}
    \tilde{C}_{\ell\ell}^{+-}(\omega)=e^{-\beta(\Delta\mu - \hbar\Omega + \hbar\omega)}\tilde{C}_{\ell\ell}^{-+}(\omega).
\end{equation}
Taking the zero-frequency limit ($\omega \to 0$) and performing the summation over all spatial coordinates $\ell$ along the two-dimensional interface area, we directly recover the result presented in Eq.~\ref{eq:imchitilde}.

\section{Analytical Evaluation of Interfacial Conductance Integrals}
\label{appendix2}

By substituting Eq.~\ref{eq:Dtildeij} into the energy integral of Eq.~\ref{eq:gLud} and performing straightforward algebraic manipulations, we obtain the master integration relation:
\begin{align}
    \label{eq:IF1F2}
    \Theta(T)\int_{-\infty}^{\infty} (\varepsilon-\Delta\mu) n(\varepsilon-\Delta\mu) \tilde{D}^{-+}(\varepsilon)\,\dd\varepsilon = \pi\hbar \Biggl[ 2\tilde{S}^2\nonumber\\
    -\frac{\tilde{S}}{\sqrt{\rho}} F_1(\lambda) + \frac{1}{2 zJ\rho\Delta\mu}F_2(\lambda,\Delta\mu,T) \Biggr] \frac{\Delta\mu\Theta(T)}{1-e^{-\beta\Delta\mu}},
\end{align}
where the functions $F_1(\lambda)$ and $F_2(\lambda,\Delta\mu,T)$ are defined as in the main text. For experimentally relevant transport parameters, the spin accumulation is on the order of $\Delta\mu \approx \hbar\Omega \sim 10^{-6}\,\text{eV}$, while the thermal energy scale $k_B T$, which is comparable to the exchange coupling $J$, is on the order of $10^{-2}\,\text{eV}$. Under these realistic operating conditions, the factor $\Delta\mu\Theta(T)(1-e^{-\beta\Delta\mu})^{-1}$ safely approaches unity.

In the continuum limit, $F_1(\lambda)$ is expressed as an integral over the first Brillouin zone. It is customary in spin-wave theory to work in the long-wavelength regime and employ the approximation $\gamma_q \approx 1 - q^2/z$; however, for the present structural integral, it is crucial to account for contributions from the entire Brillouin zone rather than restricting the analysis to the vicinity of the $\Gamma$ point ($q = 0$). To this end, we employ the binomial expansions
\begin{align}
    (1-\lambda\gamma_q)^{1/2}=\sum_{n=0}^\infty \binom{1/2}{n}(-\lambda\gamma_q)^n,
\end{align}
and
\begin{align}
    (1-\gamma_q)^{-1/2}=\sum_{m=0}^\infty \frac{(2m)!}{4^m(m!)^2}\,\gamma_q^m,
\end{align}
which allow us to express $F_1(\lambda)$ as the rigorous power series $F_1(\lambda)=\sum_{n=0}^\infty c_n \lambda^n$, with the expansion coefficients given by
\begin{align}
    c_n=\binom{1/2}{n} (-1)^n\sum_{m=0}^\infty\frac{(2m)!}{4^m(m!)^2}\,M_{n+m}.
\end{align}
In the above expression, the term
\begin{align}
    M_n=\frac{a^3}{(2\pi)^3}\int_{\text{BZ}}\gamma_q^n\,\dd^3q=\Biggl.\frac{\dd^n G(\xi)}{\dd\xi^n}\Biggr|_{\xi=0}
\end{align}
denotes the $n$-th Watson moment for the simple cubic lattice, obtained directly from the generating function
\begin{align}
    G(\xi)=\frac{a^3}{(2\pi)^3}\int_{\text{BZ}}e^{\xi\gamma_q}\,\dd^3q=I_0^3(\xi/3),
\end{align}
where $I_0(\xi)$ is the modified Bessel function of the first kind. The coefficients $c_n$ decay rapidly with increasing order $n$, justifying the truncation of the series for $F_1(\lambda)$ at second order. This yields a highly accurate polynomial approximation $F_1(\lambda)\approx 1.107-0.066\,\lambda-0.028\,\lambda^2$. Compared to a full direct numerical integration of the root over the Brillouin zone, this second-order truncated polynomial incurs a relative root-mean-square error of approximately $0.58\%$.

To evaluate the fluctuation-driven function $F_2$, we expand the Bose-Einstein distribution as: 
\begin{align}
    n(\varepsilon)=\theta(\varepsilon)\sum_{p=1}^{\infty}e^{-\beta\varepsilon p}-\theta(-\varepsilon)\sum_{p=0}^\infty e^{\beta\varepsilon p},
\end{align}
where $\varepsilon=\hbar\omega_q\pm\Delta\mu$. Since high-energy contributions are exponentially suppressed by the Boltzmann factor, the linear dispersion approximation, while strictly valid only in the long-wavelength limit $a q \ll 1$—can be mathematically extended to the entire Brillouin zone for the fluctuation channel. Therefore, we employ $\hbar \omega_q \approx \hbar c q$ and $\gamma_q \approx 1 - q^2 / z$. Furthermore, to minimize errors arising in the high-temperature regime, we introduce a Debye wave-number cutoff $q_D = (6\pi^2)^{1/3}/a$, chosen such that the corresponding spherical Brillouin zone preserves the exact volume of the original lattice Brillouin zone. The summation over momenta is then evaluated in the continuum limit using spherical coordinates, with $0 \leq q \leq q_D$, $0 \leq \theta \leq \pi/2$, and $0 \leq \phi \leq 2\pi$. After straightforward integration, we obtain
\begin{align}
    F_2(\lambda,\Delta\mu,T)= \frac{z a \Delta\mu}{\pi^2 \hbar c}\Biggl[\sum_{p=1}^\infty \Biggl( \int_0^{\varepsilon_D-\Delta\mu} u e^{-\beta p u} \,\dd u\nonumber\\
    + \int_{\varepsilon_D-\Delta\mu}^{\varepsilon_D+\Delta\mu} \frac{u\Delta\mu + u^2}{2\Delta\mu} e^{-\beta p u}\,\dd u \Biggr)+ \frac{\Delta\mu^2}{12}\Biggr],
\end{align}
with $\varepsilon_D = \hbar cq_D$. By carrying out the integrals and rewriting the resulting series in terms of polylogarithmic functions of order $s$, $\text{Li}_s$, we find
\begin{align}
    F_2(\lambda,\Delta\mu,T)=&\frac{z a \Delta\mu}{\pi^2 \hbar c} \Biggl[ \mathcal{A}(\lambda,\Delta\mu,T)- \mathcal{A}(\lambda,-\Delta\mu,T) \nonumber\\
    &+ \frac{\pi^2}{6}(k_B T)^2+ \frac{\Delta\mu^2}{12}\Biggr],
\end{align}
where we define the zone-edge damping function as
\begin{align}
    &\mathcal{A}(\lambda,\Delta\mu,T)= \frac{1}{2\Delta\mu \beta^3}\left[ \beta^2 \varepsilon_D(\varepsilon_D+\Delta\mu) \text{Li}_1\!\bigl(e^{-\beta(\varepsilon_D+\Delta\mu)}\bigr)       \right.\nonumber\\
    &\left.+ \beta(2\varepsilon_D+\Delta\mu) \text{Li}_2\!\bigl(e^{-\beta(\varepsilon_D+\Delta\mu)}\bigr)+ 2\,\text{Li}_3\!\bigl(e^{-\beta(\varepsilon_D+\Delta\mu)}\bigr)\right].
\end{align}
Assuming $\Delta\mu \ll \varepsilon_D$, we expand $\mathcal{A}$ to first order in the spin accumulation $\Delta\mu$, which yields $\mathcal{A}(\lambda,\Delta\mu,T) - \mathcal{A}(\lambda,-\Delta\mu,T)\approx-\varepsilon_D^2 \operatorname{Li}_0(\zeta)-\varepsilon_D k_B T\,\operatorname{Li}_1(\zeta)-(k_B T)^2 \operatorname{Li}_2(\zeta)$, where $\zeta = e^{-\beta \varepsilon_D}$. Within the low-temperature regime under consideration, we have $\zeta\ll 1$, which enables us to closely approximate $F_2$ as
\begin{align}
    F_2(\lambda,\Delta\mu,T) \approx &\frac{z a \Delta\mu}{\pi^2 \hbar c}\Biggl[ \frac{\pi^2}{6}(k_B T)^2 - \bigl(\varepsilon_D^2 + \varepsilon_D k_B T\nonumber\\
    &+ (k_B T)^2\bigr) e^{-\beta \varepsilon_D}\Biggr].
\end{align}
In the near-zero temperature regime, the terms proportional to $e^{-\beta \varepsilon_D}$ become negligibly small, yielding a particularly simple expression for $F_2$. For temperatures approaching $T \sim T_C$, however, these exponential corrections remain active and correctly suppress any unphysical divergence of the fluctuation channels. Since $F_2(\lambda,\Delta\mu,0)=0$, we confirm the identity of this term with the fluctuation-induced spin-mixing conductance. Finally, multiplying Eq. (\ref{eq:IF1F2}) by the linear-response prefactor $\frac{\pi}{\hbar}(J_{sd}N(\epsilon_F)/aN_e)^2$ directly recovers the independent expressions for $g_{\text{cond}}(T)$ and $g_{\text{fluct}}(T)$ presented in the main text.

\bibliographystyle{apsrev4-2}
\bibliography{bibliography}
\end{document}